\documentclass[acmtog]{acmart}

\copyrightyear{2024}
\acmYear{2024}
\setcopyright{acmlicensed}\acmConference[SA Conference Papers '24]{SIGGRAPH Asia 2024 Conference Papers}{December 3--6, 2024}{Tokyo, Japan}
\acmBooktitle{SIGGRAPH Asia 2024 Conference Papers (SA Conference Papers '24), December 3--6, 2024, Tokyo, Japan}
\acmDOI{10.1145/3680528.3687703}
\acmISBN{979-8-4007-1131-2/24/12}

\usepackage{longtable}
\acmSubmissionID{1250}

\usepackage{booktabs} %

\usepackage[ruled]{algorithm2e} %
\usepackage{multicol}
\usepackage{caption}
\usepackage{bm}
\usepackage{float}
\usepackage{xcolor}

\newcommand{\revision}[2]{#2}

\SetAlFnt{\small}
\SetAlCapFnt{\small}
\SetAlCapNameFnt{\small}
\SetAlCapHSkip{0pt}

\newcommand{\dataset}{FürElise\xspace}

\begin{document}
\title{\dataset: Capturing and Physically Synthesizing Hand Motions of Piano Performance}

\author{Ruocheng Wang $^\dagger$}
\thanks{$\dagger$ These two authors contributed equally to this work.}
\orcid{0009-0005-8404-6405}
\affiliation{%
 \institution{Stanford University}
 \country{USA}
}
\author{Pei Xu$^\dagger$}
\orcid{0000-0001-7851-3971}
\affiliation{%
 \institution{Stanford University}
 \country{USA}
}
\author{Haochen Shi}
\orcid{0000-0002-3604-465X}
\affiliation{%
 \institution{Stanford University}
 \country{USA}
}
\author{Elizabeth Schumann}
\orcid{0009-0007-1331-2707}
\affiliation{%
 \institution{Stanford University}
 \country{USA}
}
\author{C. Karen Liu}
\orcid{0000-0001-5926-0905}
\affiliation{%
 \institution{Stanford University}
 \country{USA}
}

\renewcommand\shortauthors{Wang, R. et al}

\begin{abstract}
Piano playing requires agile, precise, and coordinated hand control that stretches the limits of dexterity. Hand motion models with the sophistication to accurately recreate piano playing have a wide range of applications in character animation, embodied AI, biomechanics, and VR/AR. 
In this paper, we construct a first-of-its-kind large-scale dataset that contains approximately \revision{8}{10} hours of 3D hand motion and audio from \revision{11}{15} elite-level pianists playing \revision{98}{153} pieces of classical music. To capture natural performances, we designed a markerless setup in which motions are reconstructed from multi-view videos using state-of-the-art pose estimation models. The motion data is further refined via inverse kinematics using the high-resolution MIDI key-pressing data obtained from sensors in a specialized Yamaha Disklavier piano. Leveraging the collected dataset, we developed a pipeline that \revision{generalizes and accurately simulates}{can synthesize physically-plausible} hand motions for musical scores outside of the dataset. Our approach employs a combination of imitation learning and reinforcement learning to obtain policies for physics-based bimanual control involving the interaction between hands and piano keys.
To solve the sampling efficiency problem with the large motion dataset, we use a diffusion model to generate natural reference motions, which provide high-level trajectory and fingering (finger order and placement) information. However, the generated reference motion alone does not provide sufficient accuracy for piano performance modeling.
We then further augmented the data by using musical similarity to retrieve similar motions from the captured dataset to boost the precision of the RL policy. With the proposed method, our model generates natural, dexterous motions that generalize to music from outside the training dataset.

\end{abstract}




\begin{abstract}
Piano playing requires agile, precise, and coordinated hand control that stretches the limits of dexterity. Hand motion models with the sophistication to accurately recreate piano playing have a wide range of applications in character animation, embodied AI, biomechanics, and VR/AR. 
In this paper, we construct a first-of-its-kind large-scale dataset that contains approximately \revision{8}{10} hours of 3D hand motion and audio from \revision{11}{15} elite-level pianists playing \revision{98}{153} pieces of classical music. To capture natural performances, we designed a markerless setup in which motions are reconstructed from multi-view videos using state-of-the-art pose estimation models. The motion data is further refined via inverse kinematics using the high-resolution MIDI key-pressing data obtained from sensors in a specialized Yamaha Disklavier piano. Leveraging the collected dataset, we developed a pipeline that \revision{generalizes and accurately simulates}{can synthesize physically-plausible} hand motions for musical scores outside of the dataset. Our approach employs a combination of imitation learning and reinforcement learning to obtain policies for physics-based bimanual control involving the interaction between hands and piano keys.
To solve the sampling efficiency problem with the large motion dataset, we use a diffusion model to generate natural reference motions, which provide high-level trajectory and fingering (finger order and placement) information. However, the generated reference motion alone does not provide sufficient accuracy for piano performance modeling.
We then further augmented the data by using musical similarity to retrieve similar motions from the captured dataset to boost the precision of the RL policy. With the proposed method, our model generates natural, dexterous motions that generalize to music from outside the training dataset.

\end{abstract}
\maketitle

\begin{figure}[H]
    \centering
    \includegraphics[width=\linewidth]{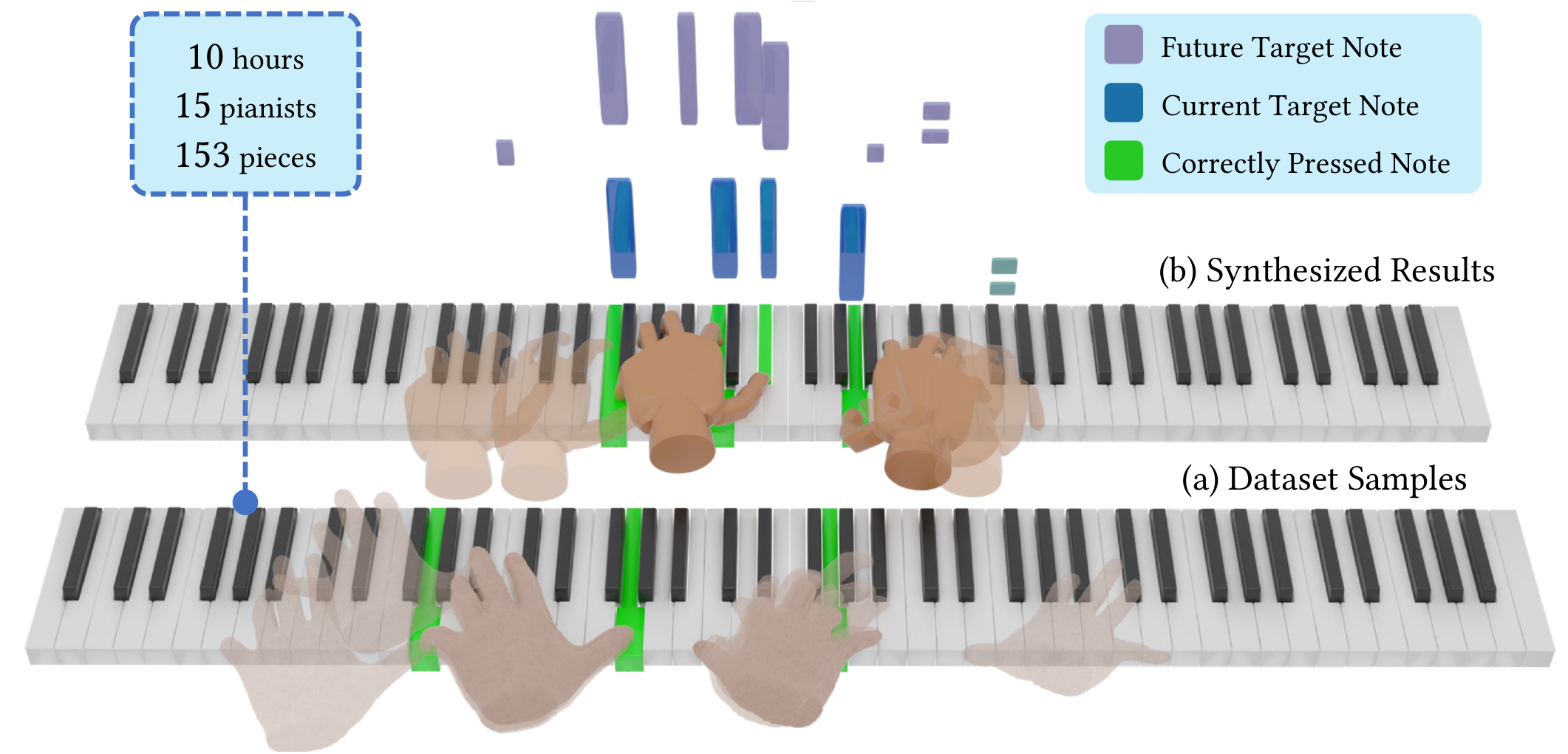}
    \caption{Our paper (a) collects the first large-scale 3D hand motion dataset of piano playing, accompanied by synchronized audio and key pressing events; (b) proposes a method that can control a physically simulated hand to play novel pieces `unheard' from the training set.} %
    \label{fig:overview}
    \vspace{-12pt}
\end{figure}

\section{Introduction}
Physically synthesizing human motion has a wide range of applications in character animation, embodied AI, AR/VR, robotics, and biomechanics. Researchers have made great strides in simulating functional and realistic human movements which enable digital agents to physically navigate and interact with environments while maintaining balance. As the application domain expands, the next frontier in human motion synthesis is to create digital agents that not only achieve motion tasks, but also exhibit elite-level athletic techniques and musical precision, comparable to the peak performance of human athletes and musicians. In this work, we take the first step toward synthesis of human peak performance through the lens of the movement of elite pianists.

Piano playing is a demanding motor skill that requires impeccable precision in finger control to press the correct keys at the correct time, agile coordination to press multiple keys simultaneously, and remarkable dexterity to fluidly play long sequences while anticipating upcoming notes. Previous works on simulating piano playing motions either rely on human-annotated fingering information (which finger to press which key)~\cite{zakka2023robopianist} or are limited to scenarios involving easier compositions~\cite{zhu2013piano, xu2022towards}.  
We believe that a better model requires a deeper understanding of how humans play the piano. 
However, there is a significant shortfall in large-scale datasets that adequately capture the diversity and complexity of piano performances. 

To address this gap, we design and build a comprehensive, non-intrusive data capture pipeline to record the 3D hand motions of pianists during their natural performances. This pipeline employs a markerless setup, where multi-view videos are processed using state-of-the-art pose estimation model~\cite{pavlakos2023reconstructing} to reconstruct 3D motions. These reconstructions are further refined through inverse kinematics, utilizing music information obtained from sensors embedded in a specialized piano~\cite{yamaha}. Using this pipeline, we have collected the first large-scale dataset of piano motions, \dataset, capturing approximately \revision{8}{10} hours of 3D hand motions from \revision{11}{15} professional or conservatory pianists performing \revision{98}{153} pieces of classical music across various genres. This dataset encompasses a broad spectrum of piano skills demonstrated by elite pianists, and includes synchronized audio, providing a valuable resource for character animation and dexterous control. \revision{}{It also enables various music-related applications such as keyboard ergonomics, music pedagogy, and pianist injury prevention.}

Leveraging \dataset, we take a step towards synthesizing physically simulated motions of piano playing for novel pieces of music ``unheard’’ from the dataset. Specifically, given a piece of sheet music, our approach first uses a diffusion model trained on the collected dataset to generate an initial reference motion that provides high-level trajectory guidance and fingering information. However, this initial reference motion often includes numerous incorrect or missing keys, making it unsuitable for training an RL policy that would ensure musically correct physical interactions between hand fingers and the piano keys. We propose to enhance this process by combining a music-based motion retrieval method with the diffusion model to create an ensemble of reference motions, thus balancing the visual performance and physical plausibility for accurate key press.
\looseness=-1

Our experiments show that, given a piece of music unseen in the training dataset, our method can synthesize natural piano motions. The policy can handle chords, fast wrist motions, and other complex piano skills, playing melodious pieces given only the sheet music. Ablations have shown that the diffusion model, music-based retrieval and reinforcement learning all contribute to the performance of the final model.

In summary, this paper makes two major contributions toward physics-based synthesis of elite-level piano performance, as illustrated in Figure~\ref{fig:overview}:
\begin{itemize}
\item We present the first large-scale dataset of 3D hand motions in piano performance with synchronized audio.
\item We develop a model that combines diffusion models, motion retrieval, and reinforcement learning to \revision{}{synthesize natural dexterous motions} playing a diverse set of piano music pieces. Our model was evaluated through extensive experiments and ablations.
\end{itemize}

\section{Related Work}

\subsection{Music2Motion}
The problem of generating motions following music has been extensively studied in recent years. \citet{Tseng2022EDGEED, alexanderson2023listen,li2021ai} 
tackle the problems of generating whole-body dancing motions from input music using diffusion models. Another line of research trains neural networks to generate upper-body motions of musicians from the audio of various instruments~\cite{Liu2020BodyMG, shlizerman2018audio, li2018skeleton, kao2020temporally,chen2021guzheng}. These works typically utilize pose estimation models to estimate 3D joint locations only from monocular videos, resulting in poor motion quality due to depth ambiguity. Moreover, these works focus on learning to generate visually plausible kinematics motions, overlooking their physical plausibility. In contrast, our work collects a large-scale high-quality dataset of piano performance motion. We propose a pipeline to train control policies that can play the novel piano pieces in a physically simulated environment. 

Apart from data-driven approaches, some early works design heuristics to animate hands for music performance. \citet{zhu2013piano} generates piano playing motion by using iterative optimizations to solve for hand trajectories that hit target keys and satisfy predefined constraints. \citet{elkoura2003handrix} considers generating left-hand motions for playing the guitar by retrieving and blending motions from a motion capture dataset. In these works, a key challenge is to determine the fingering information, which specifies which finger should press each note. Previous works rely on heuristic cost functions or additional annotations to decide fingering, which can only handle simple or manually pre-processed pieces. Our work uses a generative model trained on a large-scale dataset to provide fingering information automatically for reinforcement learning policies to learn playing unseen pieces.

\subsection{Physics-Based Dexterous Control}
Studying the control strategy for physically simulated dexterous hands has
wide applications in computer graphics, robotics, and biomechanics.
Traditional approaches usually rely on trajectory optimization and/or human-designed heuristic rules to perform control\cite{liu2008synthesis,liu2009dextrous,mordatch2012contact,wang2013video,ye2012synthesis,chen2023synthesizing}.
Most recent works on physics-based dexterous control only focus on single-hand scenarios and do not have high precision requirements\cite{andrychowicz2020learning,zhang2021manipnet,yang2022learning,xie2023hierarchical,zhao2013robust,liu2009dextrous}.
In this study, we focus on piano playing, a task that requires simultaneous bimanual control with 
exceptional temporal and spatial precision.

Piano playing is a common but intricate physical activity in daily life.
Introducing physics can help generate physically feasible motions for piano playing. Algorithms are proposed to train policies to play piano in simulations using anthropomorphic robot hands~\cite{zakka2023robopianist, xu2022towards} via reinforcement learning. Due to the complexity of the task, \citet{xu2022towards} only considers one hand playing on a simplified piano. \citet{zakka2023robopianist} leverages human annotated fingering information (which finger should press which key) to facilitate policy training. Our proposed pipeline, once trained on our large-scale dataset, can play unseen pieces without any additional annotation.

Our approach follows previous work leveraging reinforcement learning to synthesize motions under the framework of imitation learning~\cite{gail2,amp,ase,peixu2021iccgan,peixu2023composite}.
Though impressive results are achieved in generating realistic motions by imitation learning,
it is still a challenging problem to perform learning efficiently from a very large set of reference motions.
To better utilize our collected large set of piano-playing motions,
we address the problem by developing a hybrid approach to generate and retrieve motions for the policy to synthesize.
\looseness=-1

\begin{figure*}[h]
    \centering
    \includegraphics[width=\linewidth]{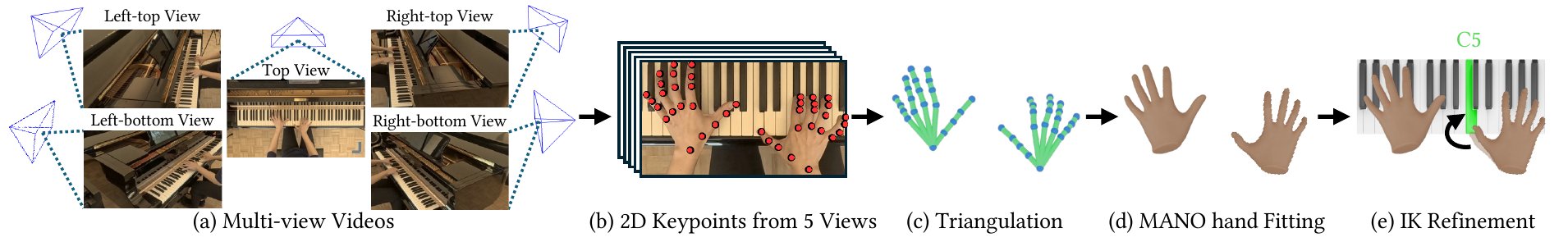}
    \caption{Overview of our pipeline to reconstruct motion data from multi-view videos. We (a) shoot 4K videos from 5 different views at 59.94 FPS using RGB camera; (b) detect 2D keypoints of the hands from each view; (c) triangulate the 2D keypoints into 3D hand skeletons with calibrated camera intrinsics and extrinsics; (d) fit the skeleton onto MANO hand meshes~\cite{manohand}; and (e) run IK with ground-truth MIDI as end effector goals to refine the finger placements for correct key pressing.}
    \label{fig:reconstruction}
\end{figure*}

\subsection{Hand Motion Datasets} Various hand motion datasets are collected in different scenarios such as grasping~\cite{taheri2020grab, chao2021dexycb}, object manipulation~\cite{fan2023arctic,wang2024dexcap}, two-hand interactions~\cite{moon2020interhand2}. However, few datasets capture the hand motions of piano performance, which are more complex and dynamics. Some works~\cite{simon2017hand, Grauman2023EgoExo4DUS} provide piano playing hand motions reconstructed using pose estimation models, but the pieces played are limited and there is no audio information recorded, which constrains their applications for tasks like MIDI-conditioned motion generation and data refinement.    \citet{wu2023marker} uses OptiTrack to reconstruct 3D hand motions of piano playing along with audio recorded in the format of Musical Instrument Digital Interface (MIDI). However, they only collected 11 pieces of music with limited variations. Our dataset contains \revision{8}{10} hours of 3D motions from \revision{}{15} elite pianists playing \revision{98}{153} different classical compositions, which cover a wide range of piano skills. All motions are provided along with MIDI audio accurately recorded by the piano's built-in recorder.

\begin{figure}[t]
    \centering
    \includegraphics[width=1\linewidth]{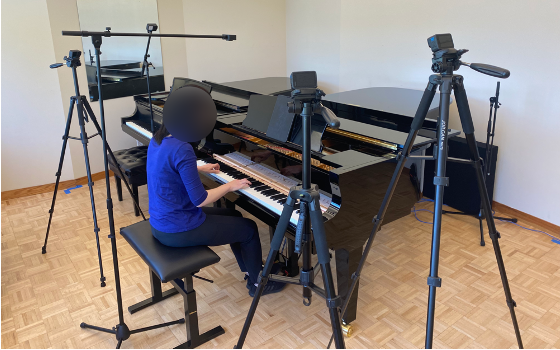}
    \caption{Data capture setup. Five GoPro cameras are placed around the piano to provide multi-view recordings of elite pianists' performances.}
    \label{fig:capture_setup}
    \vspace{-12pt}
\end{figure}

\begin{figure*}[t]
    \centering
    \includegraphics[width=\linewidth]{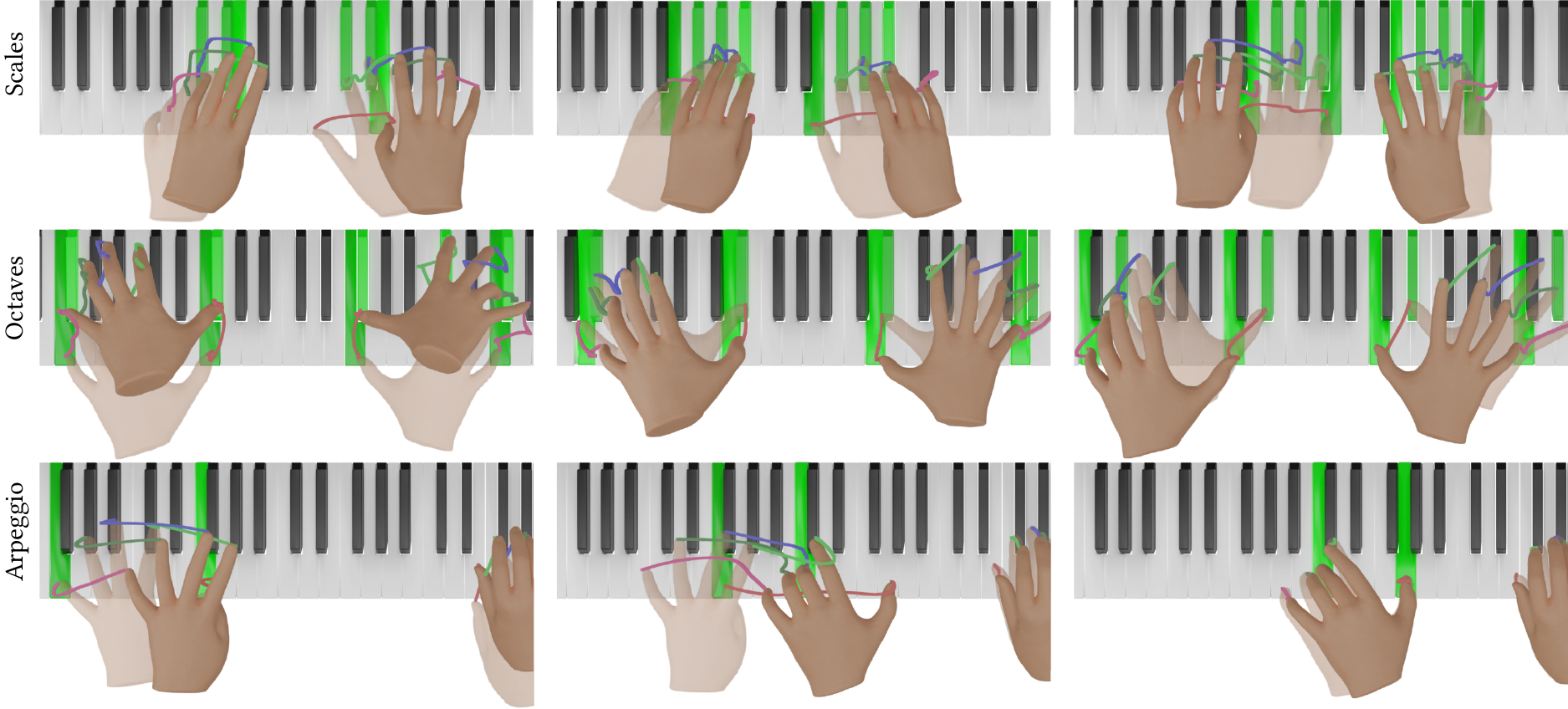}
    \caption{Examples of some piano skills in our dataset, including scales, octaves, and arpeggio. The trajectory of each fingertip is visualized. The green keys show the pressed keys through the trajectory.}
    \label{fig:dataset_examples}
\end{figure*}
\section{Dataset}
To study hand motions during piano playing, we collect a large-scale dataset, \dataset,  with approximately \revision{8.3}{10} hours of 3D hand motions paired with synchronized audio. In this section, we will first elaborate on the data capture and processing pipeline and provide an analysis of the dataset.

\subsection{Data Capture}
We aim to collect a large-scale dataset of piano playing motion performed by professional and conservatory-level pianists with minimal intrusion. 
\paragraph{Device Setup.} We record the data in a typical piano studio familiar to the performers, as shown in Figure~\ref{fig:capture_setup}. To minimize the influence of capture device, we design a markerless setup using multiview RGB cameras. Five calibrated GoPro cameras are placed around a grand piano to record synchronized videos and audio with $59.94$ FPS. All the videos have a resolution of $3840\times2160$. The grand piano is a Yamaha Disklavier DS7X ENPRO, which has a built-in recorder to record the key and pedal pressing events during the performance with high precision in MIDI format, from which the original audio with high fidelity can be reproduced. 
\paragraph{Vision-based Motion Reconstruction.} Figure \ref{fig:reconstruction} summarizes the motion reconstruction process. We first use the state-of-the-art pose estimation model HaMeR \cite{pavlakos2023reconstructing} to predict the hand pose $\bm{K}_\textrm{2D}\in \mathbb{R}^{N\times 5\times2\times21\times2}$, which are the 2D locations of 21 joints on each hand from all 5 camera views for a sequence of $N$ frames. While HaMeR can generate 3D meshes of MANO hands~\cite{manohand} in the camera space, we found that the predicted depths are not usable due to severe inaccuracy. As such, we only leverage the projected 2D keypoints from HaMeR and compute 3D locations of each joint $\bm{K}_{3D}\in \mathbb{R}^{N\times2\times 21\times3}$ via triangulation. RANSAC is used to filter out occluded keypoints, while a Butterworth filter is applied to every joint to enhance temporal smoothness, since HaMeR only considers one frame at a time. Next, we fit MANO hand parameters $\Theta = \{\theta, \beta, t\}$ to obtain 3D hand meshes for every frame, where $\theta\in\mathbb{R}^{N\times2\times16\times3}$, $\beta\in\mathbb{R}^{2\times45}$, $t\in t^{N\times2\times3}$ are the joint rotations, shape parameters and global translations of the two hands. The shape parameters are computed with extra hand calibration videos. Other parameters are optimized by minimizing the mean-squared error between the triangulated joint locations and MANO hand joint locations. 
\looseness=-1
\paragraph{MIDI-based Motion Refinement.} 
Vision-based motion reconstruction achieves reasonable results, but visible artifacts such as incorrect key-pressing or missing keys are quite common in the reconstructed motion. To improve the quality, we utilize the key-press information stored in the accompanying MIDI file. For each note played during the session, the MIDI file records the precise moments each key is fully pressed down and released. By assuming that the fingertip remains in contact with the key throughout the duration of the note, we can infer the positions of fingertips based on the states of the keys. Therefore, we apply inverse kinematics to ensure two key properties of the reconstructed motion: a) when a key is being pressed according to the MIDI file, at least one fingertip must be on the top surface of %
that key and have a depth below a preset threshold 
to trigger sound; b) when a key is not pressed according to the MIDI file, no fingertip should press the key deep enough to trigger the note; To prevent large modifications by IK, we only optimize the local joint rotations and the wrist orientation of each hand. We also limit the maximum change of fingertips to 1cm. A smoothness term is added to prevent abrupt changes between frames. Further details can be found in the appendix.

\subsection{Dataset Analysis} 
\paragraph{Data statistics.} We collect and reconstruct a total of \revision{8.3}{10} hours of 3D hand motions paired with synchronized MIDI. \revision{}{8} male and \revision{}{7} female elite pianists contribute a total of \revision{98}{153} classical compositions in various genres. 

\paragraph{Quality Evaluation.} Following \citet{zakka2023robopianist}, we use precision, recall and F1 to quantitatively evaluate the quality of our reconstructed motions according to the recorded MIDI:
\[\mathrm{Precision_i}=\frac{TP_i}{FP_i + TP_i}\quad \mathrm{Recall}=\frac{TP_i}{TP_i + FN_i}\]
\begin{equation}
\mathrm{F1_i}=\frac{2\mathrm{Precision_i} \cdot \mathrm{Recall_i}}{\mathrm{Precision_i} + \mathrm{Recall_i}},
\end{equation}
where $\mathrm{TP_i}$ computes the number of keys that are correctly pressed, $\mathrm{FP_i}$ computes the number of keys that are wrongly pressed, and $\mathrm{FN_i}$ computes the number of keys that the motion failed to press. We do this for every frame $i$ and average over all the frames in the dataset. To extract the pressed keys from reconstructed motions, similar to the IK procedure mentioned earlier, any fingertip horizontally over a key and below a preset threshold is treated as pressing the key. Using this evaluation protocol, we got a precision of 88.55, a recall of 92.53, and an F1 of 86.49 on the whole dataset. We also visualize our reconstructed motion and include the audio of the extracted MIDI in the supplementary video.

\paragraph{Qualitative Examples.} To demonstrate the diversity of motions in our dataset, we show examples of various primitive piano playing skills~\cite{neuhaus2008art} in Figure~\ref{fig:dataset_examples}.

\begin{figure*}[t]
    \centering
    \includegraphics[width=\linewidth]{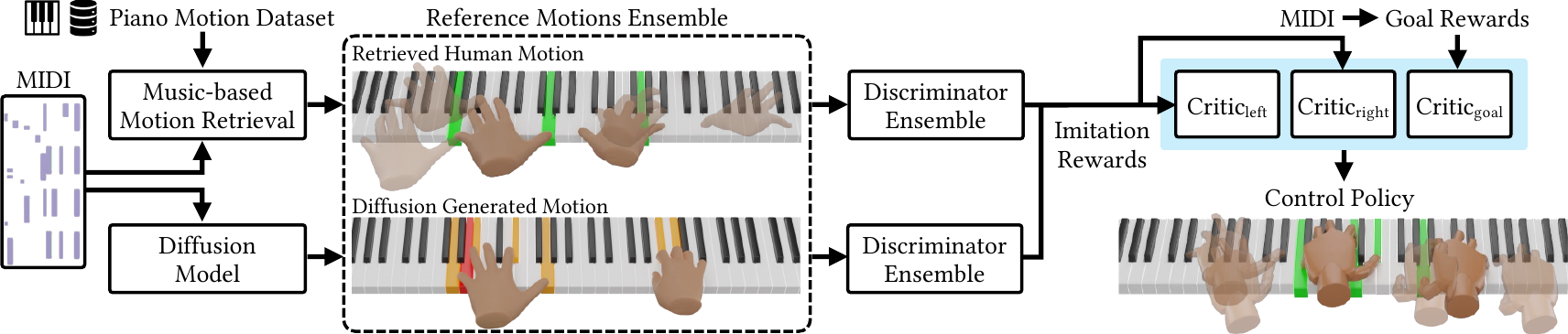}
    \caption{Overview of our method to physically simulate piano performance from a given sheet music. We use MIDI to retrieve motion data from the collected motion dataset and as input to a diffusion model for generating piano performance motions. These two sets of motions are combined into a reference motion ensemble. Utilizing the reference motions, we then employ two discriminator ensembles and three critics, which consider imitation and goal rewards, respectively, to train a control policy via reinforcement learning. }
    \label{fig:method}
    \vspace{-6pt}
\end{figure*}

\section{Play Piano with Physically Simulated Hands}
Leveraging the collected dataset, we aim to train a policy that controls two physically simulated hands in concert to play a given piece of music. Thus, the input to our method is a musical score represented as a list of notes $\{O_i = (t^{\text{start}}_i, t^{\text{end}}_i, p_i) \mid i \in \{1, 2, \ldots, n\} \}$, where $t^{\text{start}}_i, t^{\text{end}}\in\mathbb{R}$ is the start and end time of the note, and 
$p_i\in \{1, \cdots, 88\}$ %
is the pitch, which can also be mapped to one of the 88 piano keys. \revision{}{Our method finally outputs a policy that controls two hands interacting with a piano keyboard physically. A digital sound is generated by matching the pitch of the keys being pressed by the physically simulated hands.}

We propose a method that combines data-driven and physics-based approaches to achieve the goal (Figure~\ref{fig:method}). A diffusion motion model~\cite{ho2020denoising} is trained on the \dataset dataset to generate kinematics motions for the given piece. 
\revision{}{Despite the strong abilities of diffusion models to generate visually plausible motions when trained on large datasets, they often produce physically implausible artifacts, such as penetration, floating, and inconsistent interactions with objects~\cite{yuan2023physdiff,liu2024geneoh}. We observe similar issues in our settings where the generated motions exhibit seemingly plausible wrist trajectories and hand poses, but frequently exhibit incorrect contact with the keyboard, such as pressing the wrong keys or failing to press the right keys. Directly applying reinforcement learning to imitate these flawed motions would lead to unsuccessful policies. As such, we propose a music-based motion retrieval method to utilize the high-quality motions in the \dataset dataset.
Combining the diffusion-generated and retrieved motions together,}
we form an ensemble of natural \emph{or} precise reference trajectories to train an RL policy that minimizes the goal-based reward and imitates the reference motions~\cite{peixu2023composite}.

\subsection{Diffusion Model}
The goal of this module is to generate a kinematic hand trajectory given a piece of sheet music. We leverage a diffusion model, which is proven to be very effective in modeling distributions of human motions~\cite{li2023object, tevet2023human,  alexanderson2023listen, Tseng2022EDGEED} to perform kinematic motion generation.

\paragraph{Overview.} The core of the diffusion model~\cite{ho2020denoising} trains a denoiser network on dataset examples corrupted by different levels of Gaussian noises with the objective function reconstructing the original clean examples. The loss function for a conditional diffusion model is as follows:
\begin{equation}
\mathcal{L} = \mathbb{E}_{\bm{x}, t} \left[ \left\|x - \hat{\bm{x}}_\theta(\bm{x}_t, t, \bm{c}) \right\|^2 \right],
\end{equation}
where $\bm{x}$ are the clean examples, $x_t$ is the corrupted examples on noise level $t$, $\bm{c}$ is the condition vector. After training, conditional samples can be drawn by running the denoiser network iteratively on a trajectory of Gaussian noises.

\paragraph{Motion Representation.} Since the task requires high precision for the location of fingertips, we represent the dual-hand motion as a trajectory of $2\times21$ joint locations $\bm{K}\in\mathbb{R}^{M\times2\times21}$ defined in MANO hands~\cite{manohand}, similar to~\cite{liu2024geneoh}. $M$ is the number of frames considered for the diffusion model. Here we use $M=120$ which corresponds to a window of 120 frames.  To ensure consistent bone lengths during generation, we fit MANO hand models to the generated trajectory with fixed shape parameters to achieve the final predicted joint locations. 
\paragraph{Condition Representation.} To compute the condition vector $\bm{c}_T$, we first quantize input sheet music $\{O_i = (t^{\text{start}}_i, t^{\text{end}}_i, p_i) \mid i \in \{1, 2, \ldots, n\} \}$ into a binary matrix $\bm{C}\in\{0, 1\}^{N\times 88}$, where $N$ is the total number of frames in the input music 
 and $n$ is the total number of notes. Then, we divide each non-zero entry in the matrix by the duration of the corresponding key being pressed:
\begin{equation}
    \bm{C}_{i,p} = \frac{1}{t_{ip} - t^\text{start}_{ip} + 1}.
\end{equation}
In this way, the key information, as well as the duration information, are encoded into the condition vector $\mathbf{c}_T\in\mathbb{R}^{88}$.

\paragraph{Model Architecture.} We leverage a transformer-based architecture proposed in $EDGE$~\cite{Tseng2022EDGEED} to train our model. In accordance with our dataset, motions and music are quantized into 59.94FPS. The diffusion model, therefore, generates 2 seconds by outputting the results of 120 frames at a time. 

\paragraph{Long-form Generation.} Although we train the diffusion model on a window of 2 seconds, we can generate arbitrary long sequences from conditions by denoising a batch of sequences while enforcing the adjacent sequences in the batch share an overlapping path, following~\cite{Tseng2022EDGEED}.

\subsection{Music-Based Motion Retrieval}

\revision{}{To complement the diffusion-generated motions, we retrieve additional reference motions from the whole dataset for reinforcement learning policy to perform imitation learning. To do so,}
first, we quantize all notes in the dataset $\{O_i = (t^{\text{start}}_i, t^{\text{end}}_i, p_i) \mid i \in \{1, 2, \ldots, n\} \}$ into a binary matrix $\bm{M}\in\{0, 1\}^{N\times 88}$ that align with the frames of hand motions. 
We perform the same quantization for the input piece to obtain binary a matrix $\bm{M}'\in\{0, 1\}^{N'\times 88}$. Next, we compute a sliding window of length 30 and stride 1 individually over $\bm{M}$ and $\bm{M}'$ to obtain $\bm{W}\in\{0, 1\}^{N_w\times30\times88}$, $\bm{W}'\in\{0, 1\}^{N_w'\times30\times88}$. $N_w$ and $N'_w$ are the numbers of windows for the dataset and the input piece. We then compute matching from windows in the target piece to those of the dataset by minimizing their L2 distances:
\vspace{-3pt}
\begin{equation}
    \bm{c}_{j} = \arg \min_{i \in \{1, 2, \ldots, N_w\}} \|\bm{W}_i - \bm{W}'_j\|_2\quad\quad\forall j \in \{1, 2,\ldots, N_w'\}
\end{equation}
\vspace{-3pt}

This produces $N_w'$ windows of musical pieces from the dataset 
We then retrieve the corresponding hand motions, merge the overlapping windows, and generate a list of reference motions 
These motions are combined with the diffusion-generated motions to train the policy more effectively.

\subsection{Policy Training for Physics-based Control}
We set up our simulation environment using IsaacGym~\cite{makoviychuk2021isaac}.
While the simulation runs at 240 FPS,
the control runs at 60 FPS which is consistent with our diffusion model.
Our physics-based hand models are modified from~\citep{kumar2015mujoco} with geometry optimized according to the mocap subjects.
Each hand has 17 links with 27 degrees of freedom (DoFs) driven by PD servos, where the wrist has 6 DoFs, the MCP joints have 2 DoFs except that the thumb MCP has 3, and all the PIP and DIP joints have 1 DoF. 
This leads to an action space of $\mathbf{a}_t\in\mathbb{R}^{2\times27}$ for two hands.
Similar to our diffusion model,
we take the key-based binary vector as the goal representation for key pressing.
To balance the goal vector size and the observation horizon,
we utilize a compressed representation by merging 
the same key-pressing goal in consecutive frames into one.
We take the future five merged goals as the goal state with an additional timer variable that indicates the time (in terms of the number of simulation frames) left for the associated key-pressing goal. Thus the final goal state vector is of shape  $\mathbf{g}_t\in\mathbb{R}^{5\times(88+1)}$.  
To perform control,
we take a 2-frame historical observation composed of the position, orientation, and linear and angular velocities of all the links of two hands.
This results in a pose state vector $\mathbf{s}_t\in\mathbb{R}^{2\times2\times208}$ for two hands.

Due to the limited performance of the motion generated by the diffusion model,
we do not directly perform motion tracking during the control policy training.
Rather, we take the generated and retrieved motions as the reference simultaneously, and
perform imitation learning using reinforcement learning with a GAN-like architecture~\cite{peixu2021iccgan} for motion synthesis.
Following the previous literature~\cite{peixu2023composite},
to utilize the reference motions more effectively,
we decouple the motions of two hands and employ two discriminators at the same time for motion imitation of the left and right hand respectively.
By doing so, the pose of one hand does not rely on that of the other hand anymore.
We, thereby, facilitate the single-hand motion imitation by performing learning independently rather than using a dual-hand state space. 
The imitation-related reward is computed by
\begin{equation}\label{eq:dis_rew}
    r_t^{\text{imit}, h}(\mathbf{s}_t^h, \mathbf{s}_{t+1}^h) = \frac{1}{N} \sum_{n=1}^N \textsc{Clip}\left(D_n^h(\mathbf{s}_t^{h}, \mathbf{s}_{t+1}^{h}), -1, 1\right),
\end{equation}
where $h\in\{L, R\}$ indicates the imitation of the left and right hand respectively, $\mathbf{s}_t^h$ is the pose state of the single hand $h$, and $D_n^h$ is the discriminator trained using hinge loss~\cite{lim2017geometric}.

To encourage expected key-pressing behaviors,
besides imitation,
we also employ a goal-based reward function to evaluate the policy's key-pressing performance at each time step $t$.
The reward definition is different depending on the pressing condition of each key.

We assume that a key $k$ is pressed to generate sound if the pressed distance $p_k$ is greater than 90\% of that key's maximal travel distance $d_k$, which is defined using the allowed rotation range of that key.
For each target key $k$ that needs to be pressed, we have the reward term to encourage the correct key-pressing behavior:
\begin{equation} \label{eq:key_reward}
    r_{t, k}^+ = \begin{cases}
        1 & \text{if $p_k / d_k > 0.9$} \\
        \exp(||\mathbf{p}_i - \mathbf{p}_k|| + 0.01p_k/d_k) & \text{otherwise},
    \end{cases}
\end{equation}
where $\mathbf{p}_i$ is the global position of the target fingertip $i$, and $\mathbf{p}_k$ is the target position of the key.
To determine the target fingertip,
We extract fingering information based on the nearest finger to that key in the diffusion-generated motion.
The target position of the key is obtained using the surface center of a key horizontally and the 85\% position along the key's length axis vertically.

For each non-target key $\kappa$, $r_{t,\kappa}^{-}$ measures the errors of key pressing and is employed to penalize incorrect key-pressing behaviors: %
\begin{equation}
    r_{t, \kappa}^- = \begin{cases}
        p_\kappa / 0.9d_\kappa & \text{if key $\kappa$ is touched and } p_\kappa / d_\kappa > 0.1 \\
        0 & \text{otherwise}.
    \end{cases}
\end{equation}
To emulate a physical piano generating clear sound, we perform penalization even if the key is assumed not to trigger any sound virtually (i.e. $ p_\kappa / d_\kappa < 0.9$) but ignore trivial touch (i.e. $ p_\kappa / d_\kappa < 0.1$).
However, in difficult scenarios, key touching cannot be completely avoided. 
To prevent the policy from achieving a lower error of $r_{t,k}^-$ by not touching any key, an additional reward term is introduced to encourage correct key pressing behaviors even if some non-target keys are touched.

The overall goal-driven reward is defined as 
\begin{equation}
    r_t = \prod_k r_{t,k}^+ - 0.15 \sum_\kappa r_{t,\kappa}^- + 0.5 r_\mathrm{correct} - 0.05 r_\mathrm{energy},
\end{equation}
where $r_\mathrm{correct} = 1$ if all target keys are pressed correctly or $0$ otherwise,
and $r_\mathrm{energy}$ is a term measuring the energy consumption based on the average linear velocity of fingers and wrists between two frames:
\begin{equation}
    r_\mathrm{energy} =  \exp\left(-0.75\sum_{h\in\{L, R\}}\left(||\mathbf{v}_w^h|| + 0.1\sum_i ||\mathbf{v}_i^h|| \right)^2\right),
\end{equation}
where $\mathbf{v}_w^h$ is the velocity of one hand's wrist in the global space, $\mathbf{v}_i^h$ is the average velocity of each fingertip in the local system defined by its corresponding wrist joint.

The policy is trained using a multi-objective framework~\cite{peixu2023composite} to optimize
\begin{equation}
    \max \mathbb{E}_t\left[\sum_i w_i \bar{A}_{t,i} \pi(\mathbf{a}_t | \mathbf{g}_t, \mathbf{s}_t)\right],
\end{equation}
where $\bar{A}_{t,i}$ is the standardized  advantage that is estimated according to the achieved reward of each objective $i$,
and $w_i$ is an associated weight.
In our case, we have three objectives (two imitation objectives of left and right hand respectively, and one goal-driven objective).
To encourage the policy to perform expected key-pressing behaviors,
the associated weights are 0.9 for the goal-driven objective and 0.05 for each imitation objective. Please refer to the supplementary materials for the hyperparameters.

\section{Experimental Results}
We evaluate our method quantitatively on 14 pieces of music using the F1 score. We also conduct numerous ablation studies to analyze the impact of each component in our algorithm. Our dataset and method are best evaluated in the supplementary video with the audio turned on.

\subsection{Setup} 
\paragraph{Data.} We use 14 sheets of music to test our proposed pipeline. Although most recorded compositions in our dataset are classical, we include a wider range of genres including popular music, and jazz unseen during training. Because the chosen music pieces are very long with repetition, we select a clip of music from each piece and use it to train our model. The lengths of the clips are in the range from 14.4 to 28.94 seconds and 20.72 seconds on average. We do not modify the speed of the original music.

\paragraph{Metrics.} Similar to our data quality evaluation, we record the key-pressing states of model predictions and compare them with the input sheet music. Precision, recall, and F1 scores are computed for each frame and averaged over the whole piece. For diffusion-generated motions, we use the same heuristics used in data quality evaluation to extract the pressed keys: when a fingertip is below a preset depth and horizontally over a key, we treat the key as pressed. For physics-based policy, we directly query the key-pressing states from the physical simulator. %

\paragraph{Implementation Details.} For diffusion models, we train with a window of 120 frames (2 seconds). The training takes around 1 day on 2 NVIDIA A5000 GPUs. We train a single diffusion model for all the testing compositions. 
Policy trained with reinforcement learning takes around 1-3 days depending on the difficulty of studied music pieces on a single A5000 GPU and consumes about $2\times10^8$ to $4\times10^8$ training samples.

\subsection{Diffusion Generated Motions} We first show qualitative results in Figure~\ref{fig:rl_examples}. The diffusion models can generate natural and plausible kinematic trajectories on unseen pieces if viewed from a top-down perspective. However, the model cannot press keys accurately. The generated motions frequently float above the keys without pressing them or press the wrong keys, as shown in Figure~\ref{fig:fingertip_comparison}. These observations resonate with other works using diffusion models on whole-body motion and hand-object interactions~\cite{yuan2023physdiff,liu2024geneoh}. The observations are also supported by the quantitative results in Figure~\ref{fig:f1}.  More visualizations are shown in the supplementary video.

\subsection{Full Pipeline} 
\begin{figure}[t]
    \centering
    \includegraphics[width=\linewidth]{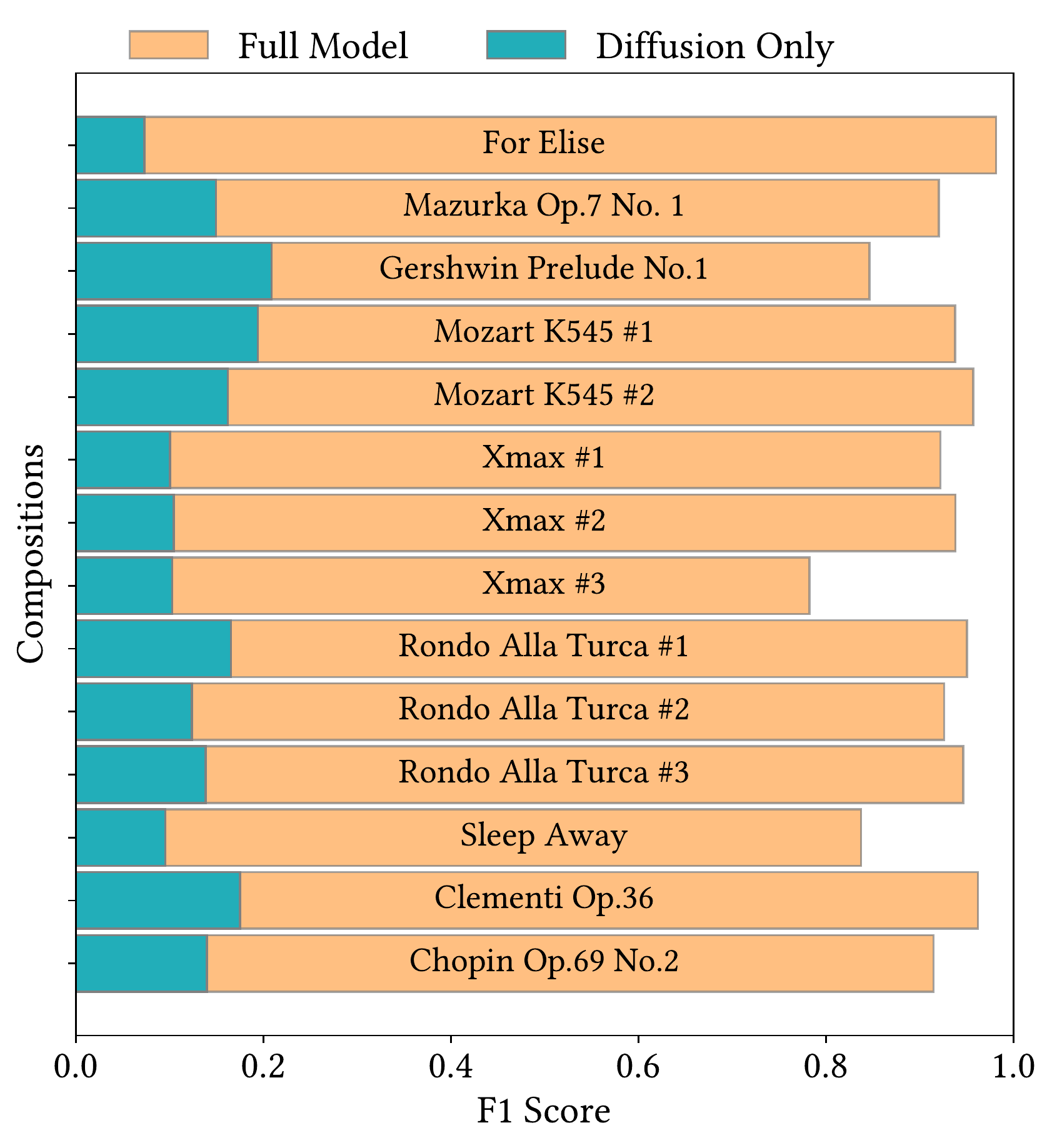}
    \vspace{-20pt}
    \caption{F1 scores of the diffusion model and our policy on the 14 test pieces. RL policies have a significant improvement over diffusion-generated motions across all 14 pieces.}
    \label{fig:f1}
\end{figure}

\begin{figure*}[t]
    \centering
    \includegraphics[width=\linewidth]{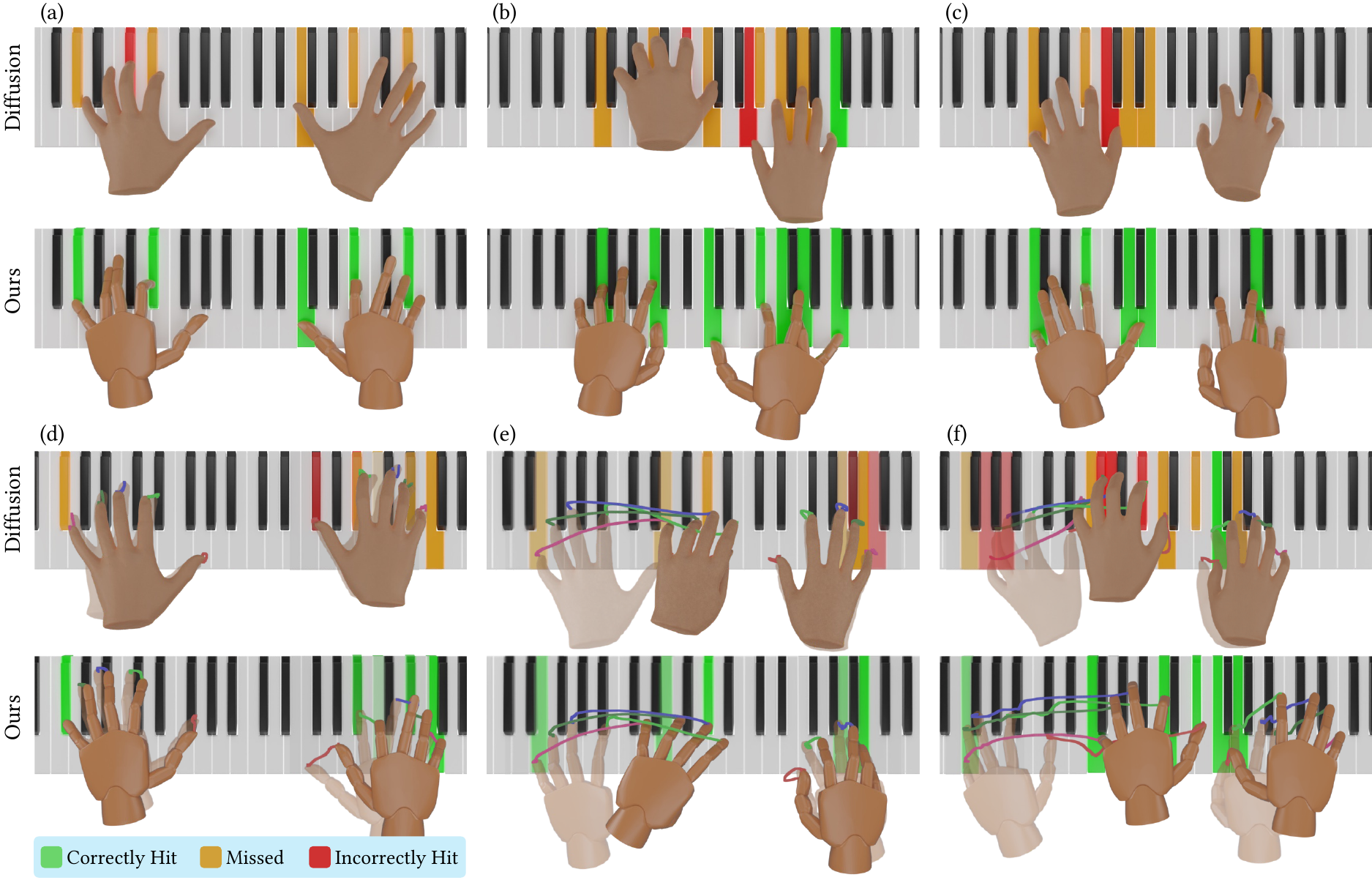}
    \caption{Results of our model accompanied with generated diffusion motions. The trajectory of each fingertip is visualized. The simulated hands can correctly press all the target keys while tying to follow the diffusion-generated motion.  The policy can handle chords (abc), double notes (d), and large wrist movements~(ef) naturally and accurately.}
    \label{fig:rl_examples}
\end{figure*}

Quantitative results of our full pipeline are summarized in Figure~\ref{fig:f1}: the policy outperforms the diffusion model by a large margin. As shown in Figure~\ref{fig:rl_examples}, 
the policy can handle large wrist motions (Fig~\ref{fig:rl_examples}f), chords (pressing multiple keys at the same time, Fig~\ref{fig:rl_examples}abc), double notes (pressing different pairs of notes sequentially, Fig~\ref{fig:rl_examples}d),  as well as arpeggios (pressing individual notes of a chord in sequence, Fig~\ref{fig:rl_examples}e). 
\revision{}{Despite the average F1 scores being as high as more than 0.8 for all the tested songs, the policy still could perform unexpected key pressing, though lasting for a very short duration. This sometimes leads to a negative impact on humans' auditory perception more than what F1 scores can reflect.} 

\begin{figure}
    \input{fig_texts/fingertip_comparison}
\end{figure}

\begin{figure}
\input{fig_texts/ablation_training}
\end{figure}
\begin{figure*}[t]
    \centering
    \includegraphics[width=\linewidth]{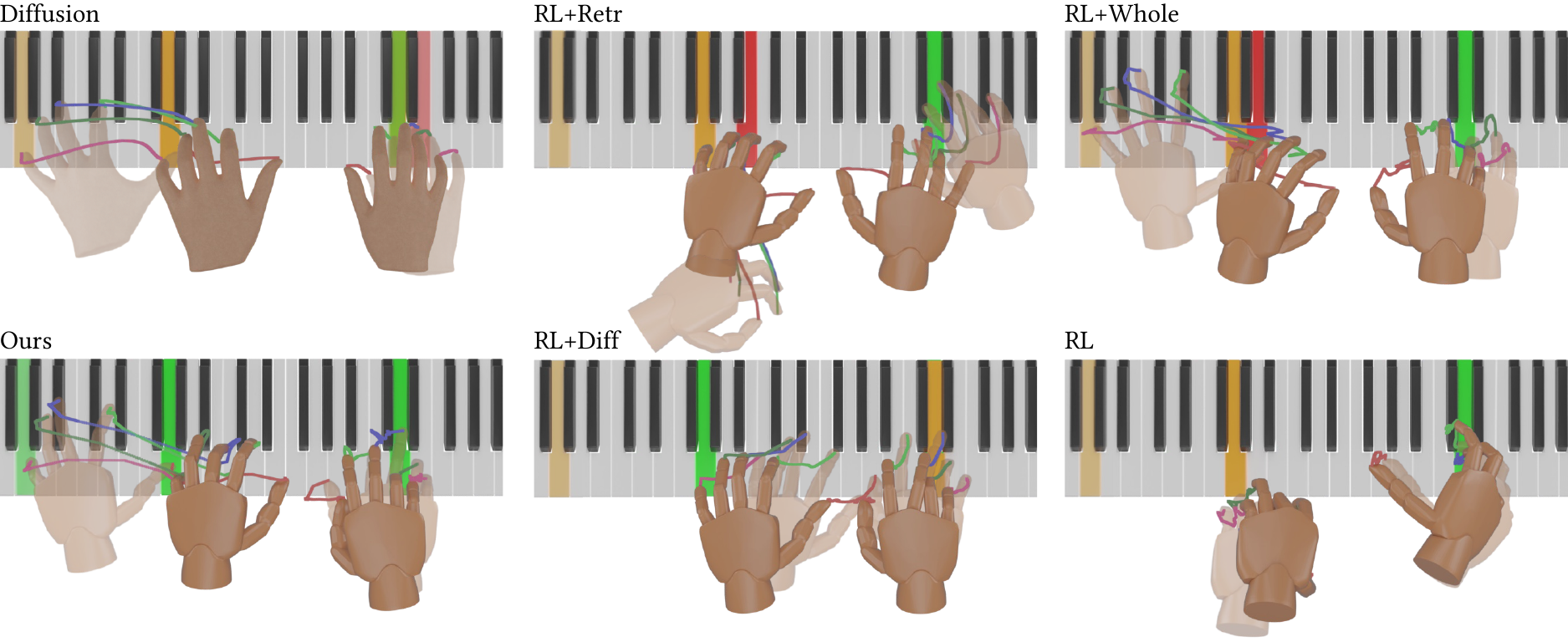}
    \caption{Comparisons between our full model and the ablative models. Without diffusion guidance, the ablative models \textit{RL}, \textit{RL+Retr}, and \textit{RL+Whole} have excessive or unnatural movements because no motions corresponding to the given piece are provided during training. Without retrieved motions from the dataset (\textit{RL+Diff}), the model tends to overfit the imprecise motions generated by the diffusion models, resulting in lower accuracy.}
    \label{fig:ablation_examples}
\end{figure*}
\subsection{Ablations}
To understand the effect of using an ensemble of motions generated by the diffusion model and those retrieved from the dataset as the reference for the control policy to learn, we design the following ablation studies tested on four music pieces: 
 
\begin{itemize}
    \item \textit{RL+Retr.} The policy is trained with only the reference motion retrieved from the dataset.

\item \textit{RL+Diff.} The policy is trained with only the reference motion generated by the diffusion model.

\item \textit{RL Only.} The policy is trained only using the goal-driven reward without motion imitation.

\item \textit{RL+Whole.} The policy is trained only using the whole motion dataset as the reference for imitation without motions generated by the diffusion model.
\end{itemize}

\paragraph{Results.}
The performance of each model is listed in Table~\ref{tbl:ablation}. The training curve is shown in Figure~\ref{fig:ablation_training}.
The full model outperforms the ablative models by a large margin in all the tested cases. We show qualitative comparisons visually of the studied ablative models in Figure~\ref{fig:ablation_examples}. 
{As we can see, the \textit{RL} only case performs the worst and behaves in a manner not human-like, which highlights the necessity of using motion imitation to ensure the motion naturalness and to help better key-pressing task execution.}
When the policies are trained without diffusion-generated motions (\textit{RL+Retr} and \textit{RL+Whole}), they yield unnatural hand poses due to the lack of fingering information. The policies also tend to have redundant motions during playing in this case because during training they could try to imitate some unrelated motions that may not strictly apply to the input music piece. When the model is only trained with diffusion-generated motions (\textit{RL+Diff}), the policy tends to overfit the erroneous finger placements existing in the diffusion-generated motions and thus has lower accuracy of key pressing. Those results demonstrate that the diffusion model and motion retrieval are complimentary
 and both of them are crucial to the final performance of our pipeline. \revision{}{Additionally, in supplementary materials, we qualitatively compare the motions generated by our control policies to those in our dataset when facing the same target notes.}
 
\begin{table}[t]
\centering
\footnotesize
\caption{We study 4 ablations of our method on 4 different pieces. We show that the F1 score of our method is significantly higher than all the variants.}
\begin{tabular}{lccccc}
\toprule
           & Ours & RL+Retr & RL+Diff& RL+Whole & RL\\
\midrule

For Elise  & \textbf{98.15} & 78.17 & 85.77 & 80.37 & 73.20 \\
Rondo Alla Turca \#3 &\textbf{94.65} & 81.94 & 78.73 & 88.51 & 34.36 \\
Clementi Op.36  & \textbf{96.21} & 95.00 & 94.17 & 79.10 & 75.28 \\
Sleep Away  &\textbf{83.75} & 53.33 & 64.13 & 49.38 & 49.97 \\
\bottomrule
\end{tabular}
\vspace{-12pt}
\label{tbl:ablation}
\end{table}

\section{Conclusion}
We present a first-of-its-kind large-scale dataset of 3D hand motion and audio of piano performance. Our dataset, \dataset, contains 8 hours of performance from 11 elite-level pianists playing 98 pieces of classical music. Leveraging \dataset, we propose a physics-based method to synthesize accurate piano playing motion for music outside the training dataset. We evaluate our method through extensive experiments and ablations.

Our work takes the first step toward motion synthesis of human peak performance using data collected from musicians for unseen songs. However, there is still a significant gap between the skill level our model achieves and that of human pianists. Several limitations in our current work might contribute to this gap. First, our method does not consider sound amplitude, a critical element in music performance. Consequently, our current model generates music with constant amplitude. 
However, the key-pressing velocity, which determines amplitude, is recorded in our dataset and can be utilized for future work. 
Second, we let the model determine fingering,
resulting in policies that may struggle with some basic skills such as finger crossover. Future work could incorporate high-level, common fingering rules to facilitate policy learning. 
\revision{}{Moreover, we leverage F1 scores to evaluate performance averaged over each frame, which may not align well with humans' auditory perception, as humans could be sensitive to some transient errors that contribute little to F1 scores such as breaking a chord or inconsistent tempo. Developing a better audio evaluation metric that meets humans' perceptions would be a great direction for future work.}
Finally, while the simulated hand models have a reasonably accurate kinematic structure, they can exert unnaturally large joint torques or generate infeasible acceleration. A promising future direction is to consider a realistic hand musculoskeletal model that generates motion through muscle activation, providing a computational tool for biomechanics studies and injury prevention.
\looseness=-1

\begin{acks}
We thank Yifeng Jiang and Jiaman Li for providing detailed feedback on the paper. This work was supported in part by the Wu-Tsai Human Performance Alliances, Stanford Institute for Human-Centered Artificial Intelligence and Roblox. We thank the 15 pianist volunteers for their essential contributions to this study. To protect their privacy, they remain unnamed, but their participation was invaluable to our research.

\end{acks}

\bibliographystyle{ACM-Reference-Format}
\bibliography{reference}

\clearpage
\appendix
\renewcommand{\thefigure}{S\arabic{figure}}
\renewcommand{\thetable}{S\arabic{table}}
\setcounter{figure}{0}
\setcounter{table}{0}
\setcounter{equation}{0}
\setcounter{page}{1}

\section{Data Capture Details}
 
We refine our reconstructed motions using the MIDI recorded during the motion capture to obtain audio-synchronized motions. All the MIDI files were recorded by the piano's built-in recorder with very high accuracy.

\subsection{MIDI synchronization} Since the MIDI and each video are recorded separately by the piano and cameras, we perform a synchronization procedure to align them temporally. We first use \citet{Kong2020HighResolutionPT} to transcribe the audio of the video to MIDI format. Then, we iterate a list of candidate offsets and find the offset where the audio and the MIDI have the maximum number of notes matched. Two notes $(t^{\text{start}}_0, t^{\text{end}}_0, p_0)$ and $(t^{\text{start}}_1, t^{\text{end}}_1, p_1)$ are treated to be matched if they have the same pitch and $\|t^{\text{start}}_0 - t^{\text{start}}_1\| \leq 0.016$. We then manually fine-tune the offset by aligning the pressing motions of fingers and the start time of the corresponding note.
\subsection{MIDI-based Inverse Kinematics.} 
To improve the quality of the reconstructed motions,
we perform inverse kinematics (IK) based on the key-pressing information provided by the MIDI files.
We first compute the pressed keys of the motions by the following heuristics:  when any fingertip is horizontally over a key and its depth is below a preset threshold, we treat that key as being pressed. We then compare the extracted pressed keys with the key-pressing information from the recorded MIDI. Frames, where the extracted pressed keys are different from the ground-truth MIDI, are considered inaccurate and the corresponding hand poses will be fixed by IK. 
We consider two possible cases of inaccurately reconstructed hand poses: (1) a muted key is wrongly pressed by any finger and (2) an activated key is omitted by all fingers for pressing. 

\paragraph{Wrongly pressed keys} For wrongly pressed keys, when multiple fingertips are pressing it, we select the fingertip with the lowest depth as the IK subject. The IK target is set such that the culprit's fingertip will move out of the key at a minimum distance. 

\paragraph{Omitted keys.} For keys that all fingertips fail to press, we first find the fingertip closest to the key by projecting all fingertips onto the surface of the key and assume the one with minimum distance to the projected location as the one performing pressing as well as the IK subject. We then set the IK target to the projected point. 

\vspace{\baselineskip}

We invalidate IK targets that need to move the fingertips for more than 1cm, and set up IK by minimizing the following loss function for every frame:

\begin{equation}
\mathcal{L}(\bm{\Theta^t})_{\text{ik}} = \frac{1}{\sum_{i=1}^{10} I_i^t} \sum_{i=1}^{10} I_i^t \|\bm{p}_i^t - \bm{\hat{p}}_i(\bm{\Theta}^t)\|^2,
\end{equation}

where:
\begin{itemize}
    \item $t$ is the index of the frame;
    \item $I_i^t$ is the mask for the $i$-th tip, with $I_i^t = 1$ if the tip is to be included in the IK and $I^t_i = 0$ otherwise;
    \item $\bm{p}_i^t$ is the target position of the $i$-th tip;
    \item $\bm{\hat{p}}_i(\bm{\Theta}^t)$ is the position of the $i$-th tip given the hand parameters $\bm{\Theta}$; and
    \item $\bm{\Theta}^t$ represents the hand parameters (pose and shape and translation) in the MANO model.
\end{itemize}

Since IK is performed only on frames with wrong key-pressing results, we further add a smoothing term to ensure temporal consistency:
\begin{equation}
\mathcal{L}_{\text{smooth}}(\bm{\Theta}) = \frac{1}{N-1}\sum_{t>1}^{N}\left(\|\bm{\Theta}^{t-1} - \bm{\Theta}^{t}\|_2^2\right),
\end{equation}
The final loss is computed by:
\begin{equation}
\mathcal{L} = \frac{1}{N}\sum_{i=1}^{N}\mathcal{L}(\bm{\Theta}^t)_{\text{ik}} + \lambda\mathcal{L}_{\text{smooth}},
\end{equation}
where $N$ is the total number of frames in the dataset and $\lambda=0.0005$. During optimization, we only optimize the local pose parameter and freeze other parameters, using L-BFGS~\cite{liu1989limited} optimizer iteratively for 100 epochs.

\section{Hyperparameters}
\begin{table}[t]
\centering
\caption{Hyperparameters for Model Training}
\begin{tabular}{lc}
    \toprule
    \textbf{Parameter} & \textbf{Value}\\
    \midrule
    \textit{Diffusion Model Training} \\
    \quad learning rate & 0.0004 \\
    \quad batch size & 512 \\
    \quad training epochs & 100 \\
    \midrule
    \textit{Reinforcement Learning} \\
    \quad policy network learning rate & $5 \times 10^{-6}$\\
    \quad critic network learning rate & $1 \times 10^{-4}$\\
    \quad discriminator learning rate & $1 \times 10^{-5}$\\
    \quad reward discount factor ($\gamma$) & $0.95$ \\
    \quad GAE discount factor ($\lambda$) & $0.95$ \\
    \quad surrogate clip range ($\epsilon$) & $0.2$ \\
    \quad gradient penalty coefficient ($\lambda^{GP}$) & $10$ \\
    \quad number of PPO workers (simulation instances) & $512$ \\
    \quad PPO replay buffer size & $4096$ \\
    \quad PPO batch size & $256$ \\
    \quad PPO optimization epochs & $5$ \\
    \quad discriminator replay buffer size & $8192$ \\
    \quad discriminator batch size & $512$ \\
  \bottomrule
\end{tabular}
\label{tab:hyper}
\end{table}

The hyperparameters used for diffusion model training and reinforcement learning are listed in Table~\ref{tab:hyper}.
We employ PPO~\cite{schulman2017proximal} as our backbone reinforcement learning algorithm.

\begin{figure*}[t]
    \centering
    \includegraphics[width=1\linewidth]{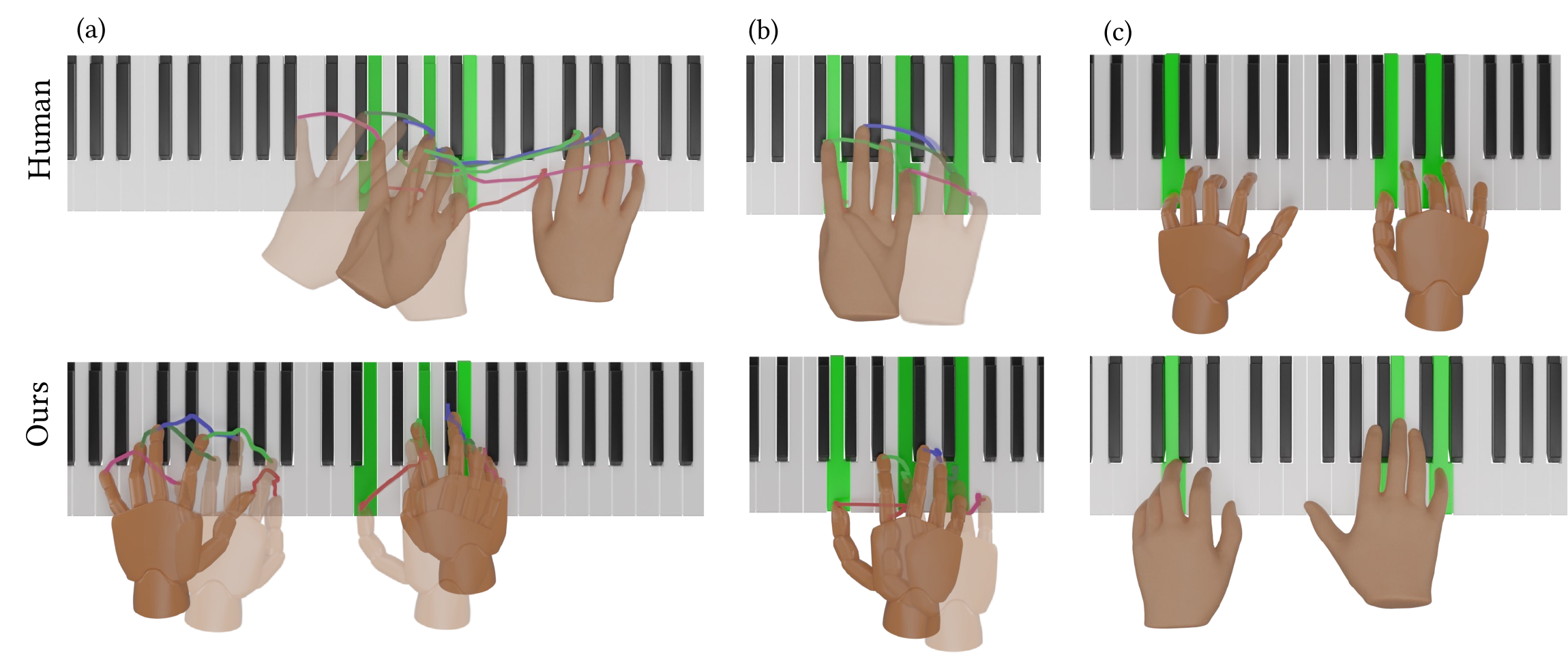}
    \vspace{-12pt}
    \caption{\revision{}{Comparison between
the synthesized motions and motions in our dataset when facing the same target notes. Our control policy could take diverse key-pressing poses by leveraging motions synthesized by the diffusion models (a), or imitating poses existing in our dataset with the similar (b) or different (c) fingering strategy.} 
}
    \label{fig:human_comparison}
\end{figure*}

\section{Additional Results}
Here, we include a qualitative comparison between
the motions generated by our control policies and those in our dataset when facing the same target notes in Figure~\ref{fig:human_comparison}. There are often multiple ways to perform the same target notes. Our pipeline enables the policy to either imitate motions generated by the diffusion model or from the captured dataset, resulting in diverse piano-playing patterns. The synthesized motions can be distinct from human pianists as shown in Figure~\ref{fig:human_comparison}a. Figure~\ref{fig:human_comparison}b shows an example where the synthesized motion largely resembles human motion in terms of fingering and hand poses. Finally, we show an example where the control policy yields results with similar hand poses but different fingering compared with the human pianist.
\section{Repertoire}
\paragraph{List of compositions in the dataset.} We include the list of all the compositions in our dataset in Table~\ref{tbl:pieces_list}.
\onecolumn
\begin{longtable}{cc}  %
\caption{List of compositions in \dataset. } \\
\label{tbl:pieces_list}\\
\vspace{-24pt} \\
\toprule
Piece & Composer \\
\midrule
Excerpt of Nocturne for the Left Hand, Op. 9, No. 2 & Alexander Scriabin \\
In meines Vaters Garten & Alma Mahler \\
Laue Sommernacht & Alma Mahler \\
Bei dir ist es traut & Alma Mahler \\
Die stille Stadt & Alma Mahler \\
Concerto No. 5 in F Major, Op. 103, "Egyptian": I. Allegro animato & Camille Saint-Saëns \\
Concerto No. 5 in F Major, Op. 103, "Egyptian": III. Molto Allegro & Camille Saint-Saëns \\
Concerto No. 5 in F Major, Op. 103, "Egyptian": II. Andante & Camille Saint-Saëns \\
Clair de Lune from Suite Bergamasque & Claude Debussy \\
Trois Chanson de Bilitis & Claude Debussy \\
Images, Book I: III. Mouvement & Claude Debussy \\
Arabesque No. 1 in E Major & Claude Debussy \\
Images, Book II: I. Cloches à travers les feuilles & Debussy \\
Images, Book II: III. Poissons d'or & Debussy \\
Images, Book II: II. Et la lune descend sur le temple qui fut & Debussy \\
Lyric Pieces, Op. 71: No. 2 Sommerabend & Edvard Grieg \\
Lyric Pieces, Op. 43: No. 1 Schmetterling & Edvard Grieg \\
Lyric Pieces, Op. 62: No. 6 Heimwärts & Edvard Grieg \\
Lyric Pieces, Op. 62: No. 4 Bächlein & Edvard Grieg \\
Lyric Pieces, Op. 54: No.3 Zug der Zwerge & Edvard Grieg \\
Lyric Pieces, Op. 38: No. 1 Berceuse & Edvard Grieg \\
Piano Quintet No. 1 in C Minor, Op. 1: II. Scherzo (Allegro vivace) & Ernő Dohnányi \\
Piano Quintet No. 1 in C Minor, Op. 1: III. Adagio, quasi andante & Ernő Dohnányi \\
Piano Quintet No. 1 in C Minor, Op. 1 IV. Finale: Allegro moderato & Ernő Dohnányi \\
Piano Quintet No. 1 in C Minor, Op. 1: I. Allegro & Ernő Dohnányi \\
Excerpt of Prelude No. 7  & Federico Mompou \\
Prelude No. 8 "On a Drop of Water" & Federico Mompou \\
Clouds & Florence Price \\
Impromptu in G-Flat Major, Op. 90, No. 3, D. 899 & Franz Schubert \\
Mazurka in A Minor, Op. 59, No. 1 & Frédéric Chopin \\
Prelude in E Minor, Op. 28, No. 4, "Largo" & Frédéric Chopin \\
Prelude in D Major, Op. 28, No. 5, "Allegro molto" & Frédéric Chopin \\
Prelude in B Minor, Op. 28, No. 6, "Lento assai" & Frédéric Chopin \\
Prelude in A Major, Op. 28, No. 7, "Andantino" & Frédéric Chopin \\
Prelude in F-Sharp Minor, Op. 28, No. 8, "Molto agitato" & Frédéric Chopin \\
Prelude in E Major, Op. 28, No. 9, "Largo" & Frédéric Chopin \\
Prelude in C-Sharp Minor, Op. 28, No. 10, "Allegro molto" & Frédéric Chopin \\
Étude in A-Flat Major, Op. 25, No. 1 & Frédéric Chopin \\
Étude in F Major, Op. 10, No. 8 & Frédéric Chopin \\
Prelude in G Major, Op. 28, No. 3, "Vivace" & Frédéric Chopin \\
Ballade No. 4 in F Minor, Op. 52 & Frédéric Chopin \\
Prelude in B Major, Op. 28, No. 11, "Vivace" & Frédéric Chopin \\
Prelude in G-Sharp Minor, Op. 28, No. 12, "Presto" & Frédéric Chopin \\
Concerto No. 1 in E Minor, Op. 11: II. Romance - Larghetto & Frédéric Chopin \\
Concerto No. 1 in E Minor, Op. 11: I. Allegro maestoso & Frédéric Chopin \\
Grande Polonaise Brilliante & Frédéric Chopin \\
Andante Spianato  & Frédéric Chopin \\
Concerto No. 1 in E Minor, Op. 11: III. Rondo - Vivace & Frédéric Chopin \\
Étude in G-Sharp Minor, Op. 25, No. 6 & Frédéric Chopin \\
Prelude in A Minor, Op. 28, No. 2, "Lento" & Frédéric Chopin \\
Étude in E Major, Op. 10, No. 3 & Frédéric Chopin \\
Piano Sonata No. 2 in B-Flat Minor, Op. 35: I. Grave – Doppio movimento & Frédéric Chopin \\
Ballade No. 3 in A-Flat Major, Op. 47 & Frédéric Chopin \\
Piano Sonata No. 2 in B-Flat Minor, Op. 35: IV. Finale: Presto & Frédéric Chopin \\
Piano Sonata No. 2 in B-Flat Minor, Op. 35: III. Marche funèbre: Lento & Frédéric Chopin \\
Waltz in A-Flat Major, Op. 34, No. 1 & Frédéric Chopin \\
Prelude in D-Flat Major, Op. 28, No. 15, "Raindrop" & Frédéric Chopin \\
Waltz in D-Flat Major, Op. 64, No. 1, "Minute Waltz" & Frédéric Chopin \\
Piano Sonata No. 2 in B-Flat Minor, Op. 35: II. Scherzo & Frédéric Chopin \\
Prelude in C Major, Op. 28, No. 1, "Agitato" & Frédéric Chopin \\
Sonata No. 3 in B Minor, Op. 58: IV. Finale: Presto, non tanto & Frédéric Chopin \\
Nocturne in D-flat major, Op. 27, No. 2 & Frédéric Chopin \\
Fantaisie-Impromptu in C-Sharp Minor, Op. 66 & Frédéric Chopin \\
Air "The Harmonious Blacksmith" & George Frideric Handel \\
Suite No. 5 in E Major: III. Courante & George Frideric Handel \\
Suite No. 5 in E Major: II. Allemande & George Frideric Handel \\
Suite No. 5 in E Major: I. Prelude & George Frideric Handel \\
Rhapsody in Blue & George Gershwin \\
Preludes, Nos. 1, 2, 3 & George Gershwin \\
 Ich atmet einen Lindenduft & Gustav Mahler \\
Twenty-six Etudes (2007) Part II: No. 10 Andantino Cantabile & H. Leslie Adams \\
Concerto in D Minor, BWV 1052: I. Allegro & J.S. Bach \\
Prelude in E Major, BWV 854 & J.S. Bach \\
Goldberg Variations, BWV 988: I. Aria & J.S. Bach \\
English Suite No. 3 in g minor, BWV 808: I. Prelude & J.S. Bach \\
English Suite No. 3 in g minor, BWV 808: II. Allemande & J.S. Bach \\
English Suite No. 3 in g minor, BWV 808: III. Courante & J.S. Bach \\
English Suite No. 3 in g minor, BWV 808: IV. Sarabande & J.S. Bach \\
English Suite No. 3 in g minor, BWV 808: V. Gavotte I & J.S. Bach \\
English Suite No. 3 in g minor, BWV 808: VI. Gavotte II (ou la musette) & J.S. Bach \\
English Suite No. 3 in g minor, BWV 808: VII. Gigue & J.S. Bach \\
Prelude in C Minor, BWV 847 & J.S. Bach \\
Italian Concerto, BWV 971: I. Allegro & J.S. Bach \\
Fugue in D Minor, BWV 875 & J.S. Bach \\
Prelude in D Minor, BWV 875 & J.S. Bach \\
Concerto in D Minor, BWV 1052: III. Allegro & J.S. Bach \\
English Suite no 6 in D minor, BWV 811, Gavotte & J.S. Bach \\
Preludes and Fugue in F Minor, BWV 881 & J.S. Bach \\
Concerto in D Minor, BWV 974: II. Adagio & J.S. Bach \\
Prelude in C Sharp Major, BWV 872 & J.S. Bach \\
Summer Hue & Jennifer Higdon \\
Intermezzo in A Major, Op. 118, No. 2: Andante teneramente & Johannes Brahms \\
Intermezzo in A Minor, Op. 116, No. 2: Andante & Johannes Brahms \\
Theme and Var 1-6 from Johannes Brahms Variations & Johannes Brahms \\
Intermezzo in A Minor, Op. 116, No. 2: Andante & Johannes Brahms \\
Andantino Cantabile & Leslie Adams \\
Trois morceaux pour piano: D’un vieux jardin & Lili Boulanger \\
Trois morceaux pour piano: D’un jardin clair  & Lili Boulanger \\
Trois morceaux pour piano: Cortège & Lili Boulanger \\
Sonata in C Major, Op. 2, No. 3: I. Allegro con brio & Ludwig van Beethoven \\
Piano Trio No. 7 in B-Flat Major, Op. 11, "Gassenhauer": I. Allegro con brio & Ludwig van Beethoven \\
Piano Sonata in B-flat Major, Op. 22: I. Allegro con brio & Ludwig van Beethoven \\
Piano Sonata No. 32 in C Minor, Op. 111: I. Maestoso – Allegro con brio ed appassionato & Ludwig van Beethoven \\
Concerto No. 5 in E-Flat Major, Op. 73: I. Allegro (excerpt) & Ludwig van Beethoven \\
Piano Sonata No. 32 in C Minor, Op. 111: II. Arietta: Adagio molto semplice e cantabile & Ludwig van Beethoven \\
Troubled Waters & Margaret Bonds \\
Miroirs: III. Une Barque sur l’Océan & Maurice Ravel \\
Not Everyone Thinks That I'm Beautiful & Michael Tilson Thomas \\
Grace & Michael Tilson Thomas \\
All Blues & Miles Davis  \\
Pictures at an Exhibition, Mvt. 1: Promenade & Modest Mussorgsky \\
Pictures at an Exhibition, Mvt. 10: The Great Gate of Kiev & Modest Mussorgsky \\
What is this thing called love & Others \\
Bewitched Bothered and Bewildered & Others \\
Slow Boat to China & Others \\
Scales & Others \\
Czerny No. 1-3 from the School of Velocity  & Others \\
Hanon No. 21 \& 22 from The Virtuoso Pianist Pt II & Others \\
Scales, Arpeggios and Chords & Others \\
Scales & Others \\
Scales in 2nds, other exercises & Others \\
Fantasie in C major, Op. 17: I. Durchaus phantastisch und leidenschaftlich vorzutragen & Robert Schumann \\
Fantasie in C major, Op. 17: III. Langsam getragen. Durchweg leise zu halten & Robert Schumann \\
Piano Sonata No. 1 in F-sharp minor, Op. 11: I. Introduzione. Un poco adagio – Allegro vivace & Robert Schumann \\
Piano Sonata No. 1 in F-sharp minor, Op. 11: II. Aria & Robert Schumann \\
Fantasie in C major, Op. 17: II. Mäßig. Durchaus energisch & Robert Schumann \\
Kinderszenen, Op. 15 No.10: Fast zu ernst & Robert Schumann \\
Kinderszenen, Op. 15 No.5: Glückes genug & Robert Schumann \\
Novellette No. 1 in F Major, Op. 21 & Robert Schumann \\
Kinderszenen, Op. 15 No.1: Von fremden Ländern und Menschen & Robert Schumann \\
Kinderszenen, Op. 15 No.2: Kuriose Geschichte & Robert Schumann \\
Kinderszenen, Op. 15 No.3: Hasche-Mann & Robert Schumann \\
Kinderszenen, Op. 15 No.4: Bittendes Kind & Robert Schumann \\
Kinderszenen, Op. 15 No.6: Wichtige Begebenheit & Robert Schumann \\
Kinderszenen, Op. 15 No.7: Träumerei & Robert Schumann \\
Kinderszenen, Op. 15 No.8: Am Kamin & Robert Schumann \\
Piano Sonata No. 1 in F-sharp minor, Op. 11: IV. Finale. Allegro un poco maestoso & Robert Schumann \\
Piano Sonata No. 1 in F-sharp minor, Op. 11: III. Scherzo e Intermezzo. Allegrissimo & Robert Schumann \\
Kinderszenen, Op. 15 No.11: Fürchtenmachen & Robert Schumann \\
Kinderszenen, Op. 15 No.12: Kind im Einschlummern & Robert Schumann \\
Kinderszenen, Op. 15 No.13: Der Dichter spricht & Robert Schumann \\
Kinderszenen, Op. 15 No.9: Ritter vom Steckenpferd & Robert Schumann \\
Expanse of my Soul & Scott Ordway \\
Études-Tableaux, Op. 39: IV. Allegro assai in b minor & Sergei Rachmaninoff \\
Études-Tableaux, Op. 39: VI. Allegro in a minor & Sergei Rachmaninoff \\
Études-Tableaux, Op. 39: VII. Lento lugubre in c minor & Sergei Rachmaninoff \\
Barcarolle in G minor, Op. 10, No. 3 & Sergei Rachmaninoff \\
Going up Yonder Improvisation & Stephen Prutsman \\
Black Pearl & Stephen Prutsman \\
Chopin Freddie & Stephen Prutsman \\
Sonata in B-Flat Major, K. 333: III. Allegretto grazioso & Wolfgang Amadeus Mozart \\
Ah, vous dirai-je, maman Variations, K. 265 & Wolfgang Amadeus Mozart \\
Sonata in B-Flat Major, K. 333: I. Allegro & Wolfgang Amadeus Mozart \\
\bottomrule
\end{longtable}

\end{document}